\documentclass{MechatronikTagung}

\usepackage{ngerman,bibgerm}
\usepackage[utf8]{inputenc} 
\usepackage[T1]{fontenc}

\usepackage{amsmath} 
\usepackage{amssymb} 

\usepackage{graphicx}

\usepackage{booktabs} 
\usepackage{cite} 

\newcommand{\RN}[1]{\uppercase\expandafter{\romannumeral#1}}  

\titeldeutsch{Nichtlineares Zweispurmodell eines LKW-Sattelaufliegers mit experimenteller Validierung der lateralen und vertikalen Reifenkräfte}

\titelenglisch{Nonlinear Two-Track Model of a Semitrailer with Experimental Validation of Lateral and Vertical Tire Forces}

\autorenliste{%
Simon F. G. Ehlers,$^1$ Zygimantas Ziaukas,$^1$ Jan-Philipp Kobler,$^2$, Alexander Busch,$^1$ Tobias Ortmaier,$^1$  Mark Wielitzka$^1$\\
$^1$Leibniz Universität Hannover, Institut für Mechatronische Systeme, 30823 Garbsen, Deutschland, simon.ehlers@imes.uni-hannover.de\\
$^2$ BPW Bergische Achsen KG, 51674 Wiehl, Deutschland\\[6pt]}

\kurzfassungdeutsch{%
Im Rahmen der Automatisierung von Nutzfahrzeugen halten verstärkt Assistenzsysteme Einzug in den LKW-Bereich. Dem Sattelauflieger kommt dabei eine wichtige Rolle zu. Dieser trägt aufgrund seiner unterschiedlich möglichen Beladungszustände und dem großen Anteil an der Gesamtmasse des Zuges, vor allem im beladenen Zustand, maßgeblich zur Fahrdynamik des gesamten Zuges bei. Um eine Grundlage für die weitere Entwicklung von Assistenzsystemen zu schaffen, wird in dieser Arbeit ein Zweispurmodell eines Sattelaufliegers entwickelt, welches die Quer- und Wankdynamik umfasst. Dies ermöglicht die nachfolgende Entwicklung von Beobachtern und Filtern zur Schätzung von nicht oder nur aufwendig messbaren Fahrzeugzuständen und -parametern. Eine Offline-Identifikation der unbekannten Modellparameter wird mittels einer Partikel-Schwarm Optimierung (PSO) durchgeführt. Die Validierung des Zweispurmodells wird auf Basis von Messfahrten eines Versuchsfahrzeugs vorgenommen, wobei der Fokus auf den lateralen und vertikalen Reifenkräften des Sattelaufliegers liegt, die durch Dehnungsmessstreifen (DMS) am Versuchsfahrzeug erfasst werden. 
}

\kurzfassungenglisch{%
As part of the automation of commercial vehicles, the number of assistance systems in this field is continuously increasing. The semitrailer plays an important role for the vehicles driving dynamics due to its highly varying loads and the large propotion to the total mass of the truck-semitrailer, especially when it is fully loaded. To create a basis for further development of assistance systems for the semitrailer, this paper presents a two-track model which includes the lateral and roll dynamics of the semitrailer. This enables the future development of observer and filter-based estimation of  vehicle states and parameters, which are impossible or very difficult to measure. For offline identification of the unknown model parameters a Particle-Swarm-Optimization (PSO) algorithm will be used. The validaton of the model is based on measurements from a test vehicle. 
The focus is on the lateral and vertical tire forces of the semitrailer, which are measured at the test vehicle using strain gauges. 
}

\begin{document}

\section{Einleitung}
Automatisiert fahrende Nutzfahrzeuge bieten in Zukunft die Chance auf hohe Wirtschaftlichkeit für den Betreiber und auf
erhöhte Sicherheit im Straßenverkehr. Bereits heutzutage werden verstärkt Advanced Driver Assistance Systems (ADAS) in Nutzfahrzeugen eingesetzt \cite{trigell2017advanced}, z.B. das elektronische Stabilitäts-Programm (ESP) oder ein Spurhalteassistent. Vor allem der Sattelauflieger hat aufgrund seiner unterschiedlich möglichen Beladungszustände und dem großen Anteil an der Gesamtmasse des Zuges maßgeblichen Einfluss auf die Fahrdynamik. Die Kenntnis über die fahrdynamischen Größen des Sattelaufliegers, zum Beispiel die lateralen und vertikalen Reifenkräfte, ermöglicht die Entwicklung weiterer Assistenzsysteme zur Überwachung und Schätzung des Verschleißes von sicherheitsrelevanten Komponenten, wie z.B. der Reifen \cite{Li_tire_wear}. Des Weiteren kann auf Basis des fahrdynamischen Modells eine Aussage über die Fahrstabilität und -sicherheit von Manövern gegeben werden, was für die Pfadplanung von automatisierten Fahrzeugen von hoher Relevanz ist. \\
Für Personenkraftwagen gibt es bereits viele Ansätze zur fahrdynamischen Modellierung mit einem Zweispurmodell und anschließender Schätzung der lateralen und vertikalen Reifenkräfte mit Beobachterstrukturen oder Filtern \cite{jin2019advanced}. 
Die Entwicklung von Modellen des Sattelaufliegers ist in der Literatur bisher jedoch weitaus weniger beschrieben, vor allem in Bezug auf die Reifenkräfte. Zwar gibt es Arbeiten, die ein Modell für die Zugmaschine und den Sattelauflieger erstellen, allerdings handelt es sich dabei meistens um Einspurmodelle, die zum Teil weder mit einem detaillierten Simulationsmodell noch in einer experimentellen Untersuchung validiert wurden \cite{brock2019comparison}. Verschiedene lineare Einspurmodelle eines Sattelzuges werden in \cite{brock2019comparison}  miteinander verglichen und die Validierung durch ein detailliertes Mehrkörper-Simulationsmodell durchgeführt. Ein nichtlineares Einspurmodell wurde in \cite{Ziaukas_19} vorgestellt und die modellbasiert berechneten lateralen Reifenkräfte auf Basis von Kraftmesssensorik validiert. Da ein Einspurmodell die Reifen einer Achse zusammengefasst betrachtet, können die lateralen Reifenkräfte nicht für einzelne Reifen gesondert modelliert werden. Auch die Dynamik der vertikalen Reifenkräfte durch den Einfluss der Querbeschleunigung und Wankdynamik kann durch ein Einspurmodell nicht abgebildet werden. Unter dem Aspekt der Verschleißschätzung einzelner Komponenten (z.B. Reifen) oder der Fahrsicherheit (z.B. Umkippen des Sattelaufliegers) kann der Einsatz eines Zweispurmodells, trotz des erhöhten Modellierungs- und Rechenaufwands, sinnvoll sein. Sowohl Einspur- als auch Zweispurmodelle von mehreren Zugmaschine-Anhängerkombinationen unter Berücksichtigung eines minimalen Rechenaufwands mit einer anschließenden experimentellen Validierung des Modells werden in \cite{ghandriz2020computationally} präsentiert. In \cite{gafvert20049} wird ebenfalls ein detailliertes Zweispurmodell aufgebaut und abschließend experimentell validiert. Allerdings liegt bei beiden Arbeiten der Fokus der Validierung nicht auf den Reifenkräften sondern auf kinematischen Größen, wie z.B. Gierrate oder Querbeschleunigung. \\
Im Gegensatz dazu wird in dieser Arbeit die Validierung eines Zweispurmodells eines Sattelaufliegers auf Basis von experimentell bestimmten lateralen und vertikalen Reifenkräften durchgeführt. Dazu wird in Kapitel \ref{kap_Modell} das mathematische Modell des Sattelaufliegers als Zweispurmodell und des Zugfahrzeugs als Einspurmodell vorgestellt. Die Identifikation der Modellparameter mittels globalem Identifikationsalgorithmus wird in Kapitel \ref{kap_Identifik} behandelt. In Kapitel \ref{kap_Validierung} wird das Modellverhalten auf Basis der Messung der lateralen und vertikalen Reifenkräfte sowie der Gier- und Wankrate  unter Betrachtung des Root-Mean-Square-Errors \mbox{(RMSE)} als Fehlermaß validiert. Eine Zusammenfassung der Erkenntnisse sowie ein Ausblick auf nachfolgende Forschungsarbeiten finden sich in Kapitel \ref{kap_Zusammenfassung}.

\section{Modellierung des Sattelzuges} \label{kap_Modell}
\begin{figure}[htbp]
	\centering
	\includegraphics[height=12cm]{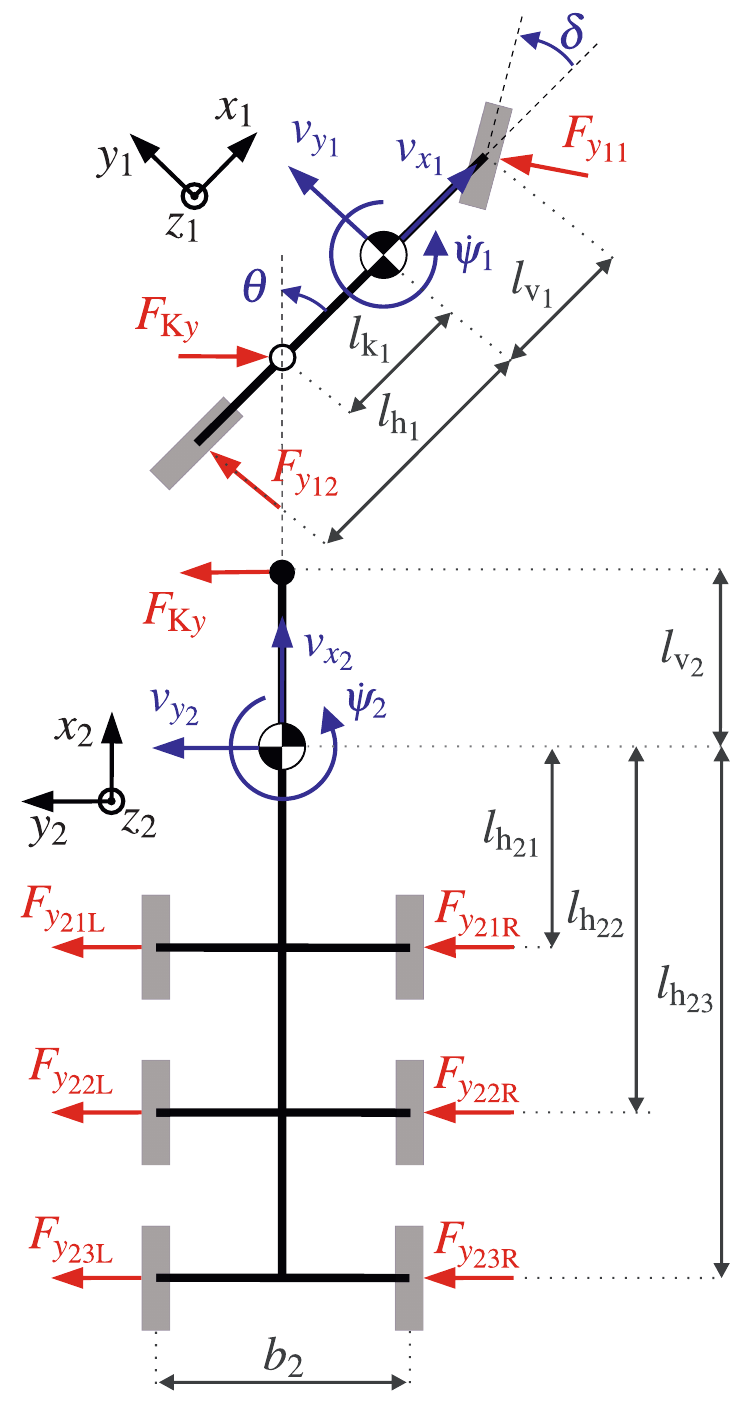}
	\caption{Freikörperbild der Zugmaschine und des Sattelaufliegers in der $x\text{-}y$-Ebene.}
	\label{fig:modell_draufsicht}
\end{figure}

In dieser Arbeit wird die Zugmaschine als Einspurmodell modelliert, da der Fokus auf dem fahrdynamischen Verhalten des Sattelaufliegers liegt. Dieses Vorgehen verringert die Modellkomplexität und den Rechenaufwand. Der Sattelauflieger hingegen ist als Zweispurmodell umgesetzt, wobei der starre Fahrzeugaufbau über Feder- und Dämpferelemente mit den sechs Rädern verbunden ist. 
Da die Kommunikation zwischen Sattelauflieger und Zugmaschine im praktischen Einsatz keine Information über Antriebs- und Bremsmomente enthält, wird hier die Modellierung der Längsdynamik nicht betrachtet. Zur Modellierung nach Newton-Euler wird das Freikörperbild in \textbf{Abbildung \ref{fig:modell_draufsicht}} verwendet. Bei der Nummerierung der Achsen steht die erste Stelle für das $i$-te Glied im Zug und die zweite Stelle für die $j$-te Achse des $i$-ten Glieds. 
Für die Zugmaschine ($i=1$) ergeben sich die Bewegungsgleichungen zu
\begin{equation} \label{eq:sum_Fy_Truck}
(4 m_{\mathrm{R}_1}+m_{\mathrm{A}_1}) a_{y_1} = F_{y_{11}} \cos(\delta) + F_{y_{12}} - F_{\mathrm{K}y} \cos(\theta) \text{,}
\end{equation}
\begin{equation} \label{eq:sum_Mz_Truck}
J_{z_1} \ddot{\psi}_1 = F_{y_{11}} \cos(\delta) l_{\mathrm{v}_1} - F_{y_{12}} l_{\mathrm{h}_1} + F_{\mathrm{K}y} \cos(\theta) l_{\mathrm{k}_1}  \text{,}
\end{equation}
wobei $a_{y_1} = \dot{v}_{y_1} + v_{x_1} \dot{\psi}_1$ die Querbeschleunigung der Zugmaschine, $F_{y_{1i}}$ die Reifenquerkräfte, $\delta$ den Lenkwinkel, $J_{z_1}$ das Massenträgheitsmoment und $\theta$ den Knickwinkel zwischen der Zugmaschine und dem Sattelauflieger beschreibt. Die Größen $l_{\mathrm{v}_1}\text{,}l_{\mathrm{h}_1}\text{ und }l_{\mathrm{k}_1}$ stellen die Abstände der Vorder- bzw. Hinterachse sowie des Koppelpunkts zum Schwerpunkt dar. Die Achs- und Radmassen sind für jedes Rad vereinfacht in $m_{\mathrm{R}_i}$ zusammengefasst. Die Aufbaumasse des jeweiligen Gliedes ist  $m_{\mathrm{A}_i}$. Die Koppelkraft $F_{\mathrm{K}y}$ wirkt nur in $y_2$-Richtung (Querrichtung) am Königszapfen des Sattelaufliegers, da die Längsdynamik vernachlässigt wird. Weitere Effekte an der Sattelkupplung, wie z.B. eine Wankabstützung, werden nicht berücksichtigt.

\begin{figure}[htbp]
	\centering
	\includegraphics[width=7cm]{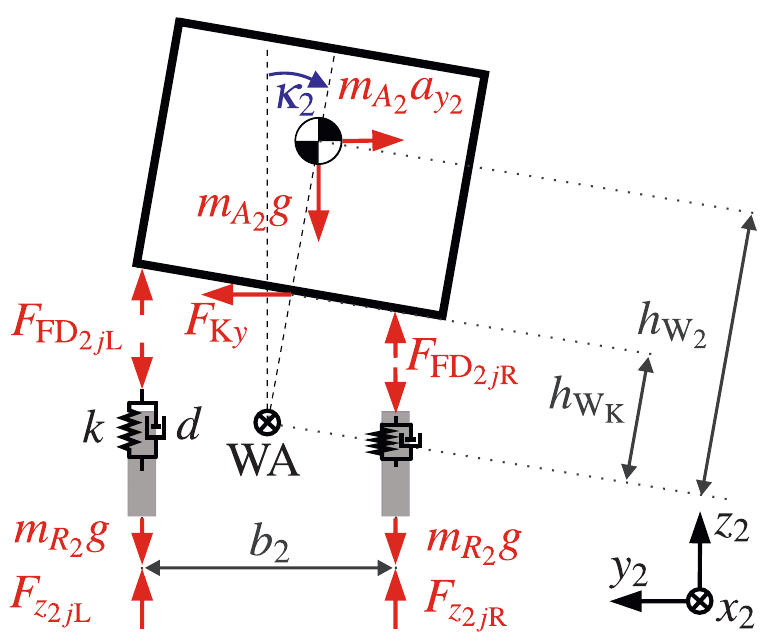}
	\caption{Freikörperbild in der  $y\text{-}z$-Ebene zur Veranschaulichung der Wankdynamik.}
	\label{fig:modell_wankdynamik}
\end{figure}

Die Bewegungsgleichungen des Sattelaufliegers ergeben sich zu
\begin{equation} \label{eq:sum_Fy_Trailer}
	(6 m_{\mathrm{R}_2}+m_{\mathrm{A}_2}) a_{y_2} + m_{\mathrm{A}_2} h_{W_2} \ddot{\kappa}_2 =  \sum F_{y_{2j\{\mathrm{L},\mathrm{R}\}}}+  F_{\mathrm{K}y} \text{,}
\end{equation}
\begin{equation} \label{eq:sum_Mz_Trailer}
\begin{split}
		J_{z_2} \ddot{\psi}_2 =  &F_{\mathrm{K}y} l_{\mathrm{v}_2}  - (F_{y_{21\mathrm{R}}} + F_{y_{21\mathrm{L}}}) l_{\mathrm{h}_{21}} \\
							     &- (F_{y_{22\mathrm{R}}} + F_{y_{22\mathrm{L}}}) l_{\mathrm{h}_{22}} - (F_{y_{23\mathrm{R}}} + F_{y_{23\mathrm{L}}}) l_{\mathrm{h}_{23}} \text{,}
\end{split}
\end{equation}
mit der Querbeschleunigung $a_{y_2} = \dot{v}_{y_2} + v_{x_2} \dot{\psi}_2$, der Wankwinkelbeschleunigung $\ddot{\kappa}_2$, dem Massenträgheitsmoment $J_{z_2}$ sowie dem Abstand des Aufbauschwerpunkts zur Wankachse $h_{W_2}$. Ein Freikörperbild der Wankdynamik ist in \textbf{Abbildung \ref{fig:modell_wankdynamik}} gezeigt. 
Das Wankverhalten wird beschrieben durch
\begin{equation} \label{eq:sum_Mx_Trailer}
	\begin{split}
	J_{x_2} \ddot{\kappa}_2 = &(a_{y_2} \cos (\kappa_2) + g \sin(\kappa_2))m_{\mathrm{A}_2} h_{\mathrm{W}_2} \\
									 &+ \frac{b_2}{2}(-F_{\mathrm{FD_{21\mathrm{R}}}}+F_{\mathrm{FD_{21\mathrm{L}}}}-F_{\mathrm{FD_{22\mathrm{R}}}} \\
									 &+F_{\mathrm{FD_{22\mathrm{L}}}}-F_{\mathrm{FD_{23\mathrm{R}}}}+F_{\mathrm{FD_{23\mathrm{L}}}}) \\
									 &-  F_{\mathrm{K}y} h_{\mathrm{W}_\mathrm{K}} \cos(\kappa_2)
	\end{split}
\end{equation}
mit der rechten bzw. linken Feder-Dämpfer Kraft $F_{\mathrm{FD_{2j\{\mathrm{L},\mathrm{R}\}}}}$ der $j$-ten Achse, der Spurweite $b_2$ und dem Abstand des Koppelpunkts zur Wankachse $h_{\mathrm{W}_\mathrm{K}}$. Das Heben des Aufbaus wird vernachlässigt.
Die Kräfte in den rechten bzw. linken Feder-Dämpfer Systemen (Steifigkeit $k$, Dämpfung $d$) der $j$-ten Achse setzen sich aus dem statischen und dynamischen Anteil der Federkraft sowie der Dämpferkraft zusammen zu
\begin{equation} 
	\begin{split}
		F_{\mathrm{FD_{2j\mathrm{R}}}} &= F_{\mathrm{stat}_{2j\mathrm{R}}} + \kappa_2 \frac{b_2}{2} k +  \dot{\kappa_2} \frac{b_2}{2} d \hspace{3mm} \text{und}\\
		F_{\mathrm{FD_{2j\mathrm{L}}}} &= F_{\mathrm{stat}_{2j\mathrm{L}}} - \kappa_2 \frac{b_2}{2} k -  \dot{\kappa_2} \frac{b_2}{2} d \text{.}
	\end{split}
\end{equation}
Zur Berechnung der statischen Aufstandskräfte an den Achsen $21$, $22$ und $23$ wird aufgrund des Luftfedersystems die Annahme getroffen, dass sich die Last auf alle drei Achsen gleichmäßig verteilt. Somit ergibt sich
\begin{equation}
	F_{\mathrm{stat}_{11}} = (m_{\mathrm{A}_1} l_{\mathrm{h}_1} e + m_{\mathrm{A}_2} l_{\mathrm{h}_{22}} (l_{\mathrm{h}_1} - l_{\mathrm{k}_1})) \frac{g}{ef} \text{,}
\end{equation}
\begin{equation}
F_{\mathrm{stat}_{12}} = (m_{\mathrm{A}_1} l_{\mathrm{v}_1} e + m_{\mathrm{A}_2} l_{\mathrm{h}_{22}} (l_{\mathrm{v}_1} + l_{\mathrm{k}_1})) \frac{g}{ef} \text{,}
\end{equation}
\begin{equation}
F_{\mathrm{stat}_{2j\{\mathrm{L},\mathrm{R}\}}} = \frac{1}{6} m_{\mathrm{A}_2} g \frac{l_{\mathrm{v}_2}}{e} \text{,}
\end{equation}
mit $e=l_{\mathrm{h}_{22}} + l_{\mathrm{v}_2}$ und $f =  l_{\mathrm{v}_1} + l_{\mathrm{h}_{1}}$. Weiterhin wird ein lineares Verhalten der Feder- und Dämpferelemente für alle Achsen angenommen. Eine weitere Annahme ist, dass sich die Reifenaufstandkräfte für die Zugmaschine aus den statischen Achskräften und den Gewichtskräften der Reifen und Achsen zusammensetzen. Für den Sattelauflieger gilt
\begin{equation}
	F_{z_{2j\{\mathrm{L},\mathrm{R}\}}} = F_{\mathrm{FD_{2j\{\mathrm{L},\mathrm{R}\}}}} + m_{\mathrm{R}_2}g \text{.}
\end{equation}
Die laterale kinematische Beziehung zwischen den beiden Gliedern des Zuges leitet sich aus der Bedingung der gleichen Geschwindigkeit beider Körper im Koppelpunkt ab
\begin{equation} \label{eq:kin_Bez}
	v_{y_1} - \dot{\psi}_1 l_{\mathrm{k}_1} = v_{x_2} \sin(\theta) + (v_{y_2} + \dot{\psi}_2 l_{\mathrm{v}_2}) \cos(\theta) \text{.}
\end{equation}
Die Winkelgeschwindigkeit des Knickwinkels setzt sich aus der Differenz der Gierraten der beiden Glieder zusammen zu
\begin{equation} \label{eq:Knickw_Gesch}
	\dot{\theta} = \dot{\psi}_2 - \dot{\psi}_1 \text{.}
\end{equation}

Zur Modellierung der statischen Reifenquerkräfte $F_{y_{\mathrm{stat},ij\{\mathrm{L},\mathrm{R}\}}}$ in Abhängigkeit des Schräglaufwinkels $\alpha$ bieten sich nichtlineare Modelle an. Vor allem bei hohen Schräglaufwinkeln, wie sie häufig beim Sattelauflieger in engen Kurven an der ersten und dritten Achse auftreten, ist ein lineares Modell nicht mehr zulässig, was in \cite{gafvert20049} anhand experimenteller Daten gezeigt wurde. In dieser Arbeit wird eine vereinfachte Version des Magic Tire Formula Model (MTFM) \cite{pacejka2006tire} genutzt
\begin{equation}
	F_{y_{\mathrm{stat},ij\{\mathrm{L},\mathrm{R}\}}} = \mu F_{z_{ij\{\mathrm{L},\mathrm{R}\}}} \sin(C_{ij}\arctan(B_{ij} \alpha_{ij\{\mathrm{L},\mathrm{R}\}}))\text{,}
\end{equation}

wie in \cite{Ziaukas_19} beschrieben.
Der Parameter $\mu$ fungiert als Skalierungsfaktor der Funktion und beschreibt unter anderem den Reibwert im Reifen-Fahrbahn-Kontakt. 
Die Funktion enthält die Koeffizienten $C_{ij}$ und 
\begin{equation}
	B_{ij} = \frac{1}{C_{ij} \mu F_{z_{ij\{\mathrm{L},\mathrm{R}\}}}} c_{1_{ij}} \sin(2 \arctan\biggl( \frac{F_{z_{ij\{\mathrm{L},\mathrm{R}\}}}}{c_{2_{ij}}}\biggr)) \text{.}
\end{equation}
Daraus folgt, dass die zur Beschreibung des Querkraftverhaltens unbekannten Parameter $\mu$, $C_{ij}$, $c_{1_{ij}}$ und $c_{2_{ij}}$ ermittelt werden müssen. Die Schräglaufwinkel ergeben sich für das Einspurmodell der Zugmaschine zu
\begin{equation}
	\begin{split}
		\alpha_{11} &= \delta - \arctan \biggl(\frac{v_{y_1}+\dot{\psi}_1 l_{\mathrm{v}_1}}{v_{x_1}}\biggr) \hspace{3mm} \text{und} \\
		\alpha_{12} &= -\arctan \biggl(\frac{v_{y_1}-\dot{\psi}_1 l_{\mathrm{h}_1}}{v_{x_1}}\biggr)\text{.}
	\end{split}
\end{equation}

Für das Zweispurmodell des Sattelaufliegers ergibt sich für die $j$-te Achse
\begin{equation}
\begin{split}
	\alpha_{2jR} &= -\arctan \biggl(\frac{v_{y_2}-\dot{\psi}_2 l_{\mathrm{h}_{2j}}+\dot{\kappa}_2 h_{W_2}}{v_{x_2}+\dot{\psi}_2 \frac{b_2}{2}}\biggr) \hspace{3mm} \text{und} \\
	\alpha_{2jL} &= -\arctan \biggl(\frac{v_{y_2}-\dot{\psi}_2 l_{\mathrm{h}_{2j}}+\dot{\kappa}_2 h_{W_2}}{v_{x_2}-\dot{\psi}_2 \frac{b_2}{2}}\biggr) \text{.}
\end{split}
\end{equation}
Die zeitliche Verzögerung des Aufbaus der lateralen Reifenkräfte bei dynamischen Manövern bzw. sich schnell ändernden Schräglaufwinkeln wird durch ein Verzögerungsglied  erster Ordnung berücksichtigt
\begin{equation} \label{eq:Verzögerung_Reifen}
	F_{y_{ij\{\mathrm{L},\mathrm{R}\}}}+\frac{l_{ij}}{v_{x_i}}\dot{F}_{y_{ij}} = F_{y_{\mathrm{stat},ij\{\mathrm{L},\mathrm{R}\}}} \text{,}
\end{equation}
wobei $l_{ij}$ die Einlauflängen sind \cite{schramm2013}. Weitere Effekte am Reifen, wie z.B. ein möglicher Stick-Slip-Effekt bei hohen Schräglaufwinkeln, werden hier nicht berücksichtigt. 

Für das Zustandsraummodell ergibt sich ein Zustandsvektor mit $n_x= 15$ Zuständen
\begin{equation}
	\begin{split}
			\pmb{x} = (&v_{y_1}, \dot{\psi}_1, \kappa_2, v_{y_2}, \dot{\kappa}_2, \dot{\psi}_2, \theta, F_{y_{11}}, F_{y_{12}},\\
			              & F_{y_{21\mathrm{R}}}, F_{y_{21\mathrm{L}}}, F_{y_{22\mathrm{R}}}, F_{y_{22\mathrm{L}}}, F_{y_{23\mathrm{R}}}, F_{y_{23\mathrm{L}}})^\mathrm{T} \text{.}
	\end{split}
\end{equation}
Um aus den vorgestellten Gleichungen ein Zustandsraummodell zu erzeugen, wird mit Gleichung (\ref{eq:sum_Fy_Truck}) die Koppelkraft $F_{\mathrm{K}y}$ eliminiert. Somit bleiben noch 4 Gleichungen (\ref{eq:sum_Mz_Truck})-(\ref{eq:sum_Mx_Trailer}) übrig. Unter Hinzunahme der zeitlichen Ableitung von (\ref{eq:kin_Bez}), der Beziehung (\ref{eq:Knickw_Gesch}) und $\dot{\pmb{x}}{(3)} = \pmb{x}{(5)}$ sowie der 8 Gleichungen der Reifenkräfte (\ref{eq:Verzögerung_Reifen}), ergeben sich insgesamt 15 Gleichungen. Diese können in die Form 
\begin{equation} \label{eq:vector_matrix_form}
	\pmb{M}\dot{\pmb{x}} = \tilde{\pmb{f}}(\pmb{x}, \pmb{u}) \hspace{3mm} \text{mit} \hspace{3mm} \pmb{u} = (\delta, v_{x_2}, \dot{v}_{x_2})^\mathrm{T}
\end{equation}
als Eingangsvektor überführt werden, wobei $\dot{v}_{x_2}$ die Längsbeschleunigung des Sattelaufliegers ist. Anschließend kann Gleichung (\ref{eq:vector_matrix_form}) zu 
\begin{equation} \label{eq:zustandsraummodell}
	\dot{\pmb{x}} = \pmb{M}^{-1} \tilde{\pmb{f}}(\pmb{x}, \pmb{u}) = \pmb{f}(\pmb{x}, \pmb{u})
\end{equation}
aufgelöst werden. 
Die $n_y=12$ Größen für die Ausgangsgleichung werden zu
\begin{equation} \label{eq:Ausgangsgleichung}
	\begin{split}
		\pmb{y} = \pmb{g}(\pmb{x}, \pmb{u}) = (&\dot{\psi}_1, \dot{\psi}_2, \dot{\kappa}_2, \theta,  F_{y_{21\mathrm{R}}}, F_{y_{21\mathrm{L}}},  F_{y_{23\mathrm{R}}}, \\
                     & F_{y_{23\mathrm{L}}}, F_{z_{21\mathrm{R}}}, F_{z_{21\mathrm{L}}}, F_{z_{23\mathrm{R}}}, F_{z_{23\mathrm{L}}})^\mathrm{T}
	\end{split}
\end{equation}
gewählt, welche im weiteren Verlauf für die Identifikation und Validierung (Kap. \ref{kap_Identifik} und \ref{kap_Validierung}) genutzt werden.

\section{Identifikation der Modellparameter} \label{kap_Identifik}
Einige der Parameter, die in dem Zustandsraummodell in (\ref{eq:zustandsraummodell}) verwendet werden, sind direkt messbar. Dazu gehören geometrische Längen, wie die Achsabstände zu dem Königszapfen von der Zugmaschine und dem Sattelauflieger. Auch die Fahrzeugmassen $m_{\mathrm{A}_1}+4m_{\mathrm{R}_1}$ bzw. $m_{\mathrm{A}_2}+6m_{\mathrm{R}_2}$ sowie die Schwerpunktpositionen in Längsrichtung der Zugmaschine $l_{\mathrm{v}_1}$ und des Sattelaufliegers $l_{\mathrm{v}_2}$ können durch Auswiegen der Achsen experimentell bestimmt werden. Die Massenträgheitsmomente $J_{z_1}$, $J_{z_2}$ und $J_{x_2}$  werden vereinfacht aus der Annahme eines Rechtecks mit homogener Massenverteilung bestimmt. Die Ersatzmassen für die Räder und Achse $m_{\mathrm{R}_1}$ und $m_{\mathrm{R}_2}$ können Datenblättern entnommen werden. 
Die weiteren Parameter werden mit einer Partikel-Schwarm-Optimierung (PSO) \cite{kennedy1995particle} bestimmt. Dazu gehören die Reifenparameter  $\mu$, $C_{ij}$, $c_{1_{ij}}$, $c_{2_{ij}}$ und $l_{ij}$. Es werden hierbei vereinfachend die gleichen Parameter für die sechs Reifen sowie für die Feder-Dämpfer-Elemente $k$ und $d$ des Sattelaufliegers angenommen, da sich die Reifen und Achsen $j = 1,2,3$ konstruktiv nicht unterscheiden. Da die Position der Wankachse $\mathrm{WA}$ sowie die Schwerpunkthöhe $h_{\mathrm{W}_2}$ des Sattelaufliegers (siehe Abb. \ref{fig:modell_wankdynamik}) experimentell schwierig zu bestimmen sind, werden diese Parameter ebenfalls offline identifiziert. Somit ergeben sich die 15 zu identifizierenden Parameter
\begin{equation}
	\begin{split}
		\pmb{p} = (&\mu, C_{11}, C_{12}, C_{2j}, c_{1_{11}}, c_{1_{12}}, c_{1_{2j}}, c_{2_{11}}, c_{2_{12}}, c_{2_{2j}},  \\
					  &l_{11}, l_{12}, l_{2j}, k, d, h_{\mathrm{W}_\mathrm{K}}, h_{\mathrm{W}_2})^\mathrm{T} \text{.}
	\end{split}
\end{equation} 
Zur Identifikation von $\pmb{p}$ wurden Messfahrten mit dem Versuchsfahrzeug (siehe Kapitel \ref{kap_Validierung}) durchgeführt, welche die Manöver Slalom, doppelter Spurwechsel und Kurvenfahrt bei jeweils unterschiedlichen Geschwindigkeiten enthalten. 
Es wird die Kostenfunktion 
\begin{equation}
	J = \sum_{l=1}^{n_y} \frac{\left\|\pmb{y}_{\mathrm{m}}(l)-\pmb{y}(l)\right\|^2}{\left\|\pmb{y}_{\mathrm{m}}(l)-\bar{\pmb{y}}_\mathrm{m} (l)\right\|^2}
\end{equation}
genutzt, wobei $\pmb{y}(l)$ der $l$-te Eintrag aus (\ref{eq:Ausgangsgleichung}) ist, $\pmb{y}_{m}(l)$ die korrespondierende gemessene Größe und $\bar{\pmb{y}}_\mathrm{m}(l)$ der jeweilige Mittelwert ist.
Ziel der PSO ist es, einen Parametersatz $\pmb{p}_{\mathrm{opt}}$ zu finden, der die Kostenfunktion $J$ minimiert.
Die Optimierung wurde auf maximal 150 Iterationen beschränkt. Auf Basis des ermittelten Parametersatzes wird anschließend eine lokale Optimierung mit sequentieller quadratischer Programmierung \cite{nocedal2006numerical} durchgeführt, was in \cite{Ziaukas_19} zu einem 3\% besseren Resultat führt. 
Da das Ergebnis der PSO stochastischen Einflüssen unterliegt, wurde die gesamte Identifikation 60 Mal wiederholt. Auf Basis aller Durchläufe wurde der Parametersatz mit dem geringsten Wert der Kostenfunktion für die experimentelle Validierung ausgewählt.

\section{Experimentelle Validierung} \label{kap_Validierung}
\subsection{Messfahrzeug}

Das Messfahrzeug besteht aus einer Zugmaschine und dem Sattelauflieger, welcher der europäischen Richtlinie 96/53/EC entspricht. 
\begin{figure}[htbp]
	\centering
	\includegraphics[width=\linewidth]{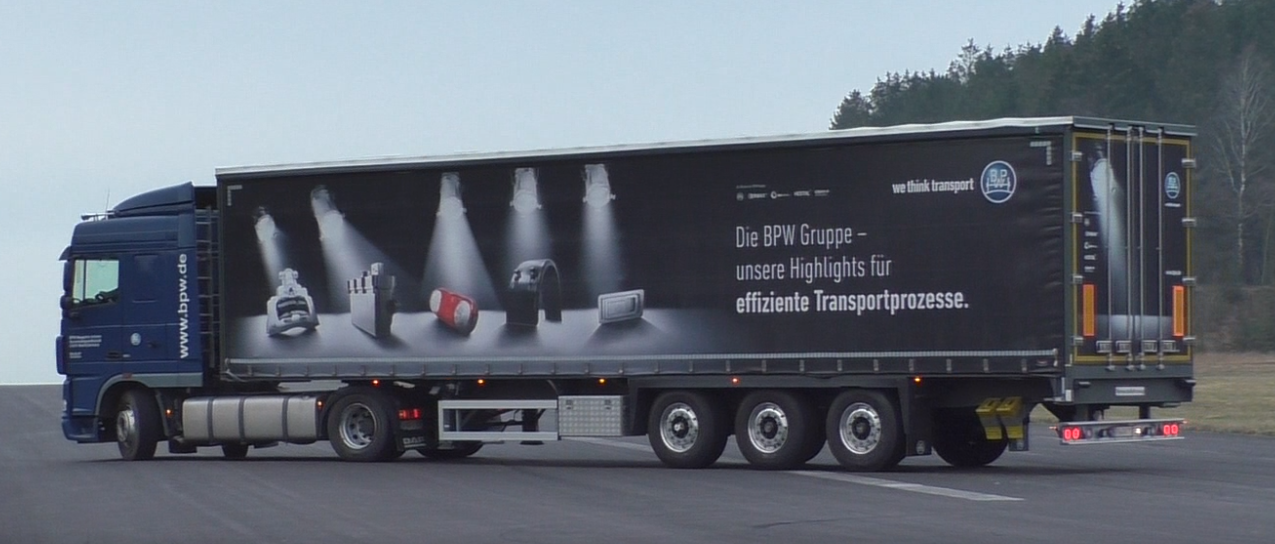}
	\caption{Im Versuch verwendete Zugmaschine-Sattelauflieger-Kombination.}
	\label{fig:versuchsfahrzeug}
\end{figure}
Der Sattelauflieger ist mit einem luftgefederten Fahrwerk (BPW Eco Air, BPW Bergische Achsen KG, Wiehl, Deutschland) sowie einem Elektronischen-Bremssystem (EBS, ZF WABCO, Friedrichshafen, Deutschland) ausgestattet. Zur Messdatenaufnahme wird ein Data-Aquisition-System (DAQ) (Dewesoft d.o.o, Trbovlje, Slovenien) verwendet, welches sowohl analoge Eingänge als auch CAN-Schnittstellen aufweist. Die Kommunikation über CAN zwischen der Zugmaschine und dem Sattelauflieger (ISO 11992) wird vom DAQ-System erfasst und beinhaltet unter anderem die Längsgeschwindigkeit des Sattelaufliegers. Des Weiteren wird der CAN-Bus der Zugmaschine (nach SAE J 1939) mit einer Frequenz von 100 Hz aufgezeichnet und beinhaltet die Gierrate $\dot{\psi}_1$ und den Lenkradwinkel $\delta$ der Zugmaschine.
Die lateralen und vertikalen Reifenkräfte $F_{y_{21\mathrm{R}}}, F_{y_{21\mathrm{L}}},  F_{y_{23\mathrm{R}}}, F_{y_{23\mathrm{L}}}, F_{z_{21\mathrm{R}}}, F_{z_{21\mathrm{L}}}, F_{z_{23\mathrm{R}}} \text{ und } F_{z_{23\mathrm{L}}}$ werden durch kalibrierte Dehnungsmessstreifen (DMS) an der ersten und dritten Achse des Sattelaufliegers gemessen und als analoge Eingänge mit 1000 Hz vom DAQ-System erfasst.  Zur Messung der kinematischen Größen (Gierrate $\dot{\psi}_2$ und Wankwinkelgeschwindigkeit $\dot{\kappa}_2$) des Sattelaufliegers wird eine 6-DOF inertiale Messeinheit bei einer Abtastrate von 100  Hz verwendet. Zur Messung des Knickwinkels $\theta$ kommt ein in den Königszapfen integrierter Winkelsensor (Abtastrate 1000 Hz) zum Einsatz. 

Zur Entwicklung des Fahrzeugmodells wurde die Software \mbox{MATLAB/Simulink} (The MathWorks, Inc., Natick, USA) und Maple (Waterloo Maple Inc., Waterloo, Kanada) genutzt. Die Identifikation der Parameter sowie die Validierung wurde nach der erfolgten Messdatenaufnahme offline in MATLAB durchgeführt.

\subsection{Validierung}

Um das Modell zu validieren, wird eine Sequenz von Manövern hinzugezogen, die nicht für die Identifikation der unbekannten Modellparameter (Kap. \ref{kap_Identifik}) genutzt wurde. Die Sequenzlänge beträgt $115 \,\textrm{s}$ und beinhaltet verschiedene querdynamische Manöver bei unterschiedlichen Geschwindigkeiten. Die Messdatenaufnahme erfolgte auf einem trockenen Asphaltuntergrund. 
\begin{figure}[htbp] 
	\centering
	\includegraphics[width=\linewidth]{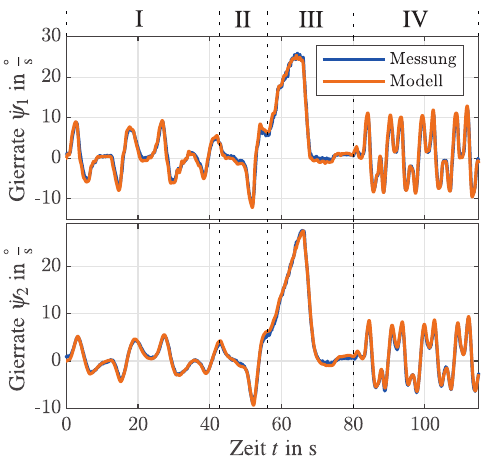}
	\caption{Gierrate von Zugmaschine und Sattelauflieger.}
	\label{fig:Gierrate}
\end{figure}
In den ersten $40\,\textrm{s}$ werden mehrere doppelte Spurwechsel bei etwa $15\,\textrm{km/h}$ gefahren (Abschnitt \RN{1}). Anschließend wird der Sattelzug auf $30\,\textrm{km/h}$ beschleunigt und ab ca. $50\,\textrm{s}$ eine Rechtskurve gefahren (Abschnitt \RN{2}). Daraufhin wird auf $18\,\textrm{km/h}$ verzögert, um eine $180^\circ$ Kehrtwende in Form einer Linkskurve auszuführen (Abschnitt \RN{3}). Nach anschließender Beschleunigung auf $40\,\textrm{km/h}$ werden erneut mehrere doppelte Spurwechsel hintereinander durchgeführt (Abschnitt \RN{4}). Somit sind in dieser Sequenz viele für den Straßenverkehr relevante Fahrsituationen abgebildet. 
\begin{figure}[htbp] 
	\centering
	\includegraphics[width=\linewidth]{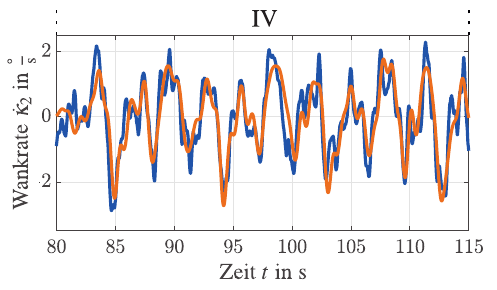}
	\caption{Wankwinkelgeschwindigkeit des Sattelaufliegers.}
	\label{fig:Wankrate}
\end{figure}

In \textbf{Abbildung \ref{fig:Gierrate}} werden die modellbasiert berechneten Gierraten $\dot{\psi}_{1,2}$ von Zugmaschine und Sattelauflieger mit den Messungen verglichen. Es ist festzuhalten, dass in allen vier Abschnitten das Modell das Gierverhalten der Glieder gut abbilden kann. Der RMSE zwischen Modell und Messung beträgt für die Gierrate der Zugmaschine $0{,}76\,\frac{^\circ}{\textrm{s}}$ und für den Sattelauflieger $0{,}39\,\frac{^\circ}{\textrm{s}}$. Zudem ist in \textbf{Abbildung \ref{fig:Wankrate}} die Wankwinkelgeschwindigkeit des Sattelaufliegers dargestellt, allerdings nur für Abschnitt \RN{4}. Da in den Abschnitten \RN{1}-\RN{3} nur eine geringe Anregung des Wankverhaltens vorhanden ist, lässt sich aufgrund des geringen Signal-Rausch-Verhältnisses der Messung keine Aussage über die Modellgüte treffen. Für den Abschnitt \RN{4} ist jedoch erkennbar, dass die Wankdynamik ausreichend angeregt wird und diese durch das Modell ebenfalls abgebildet werden kann (RMSE von $0{,}5\,\frac{^\circ}{\textrm{s}}$). 
\begin{figure}[htbp]
	\centering
	\includegraphics[width=\linewidth]{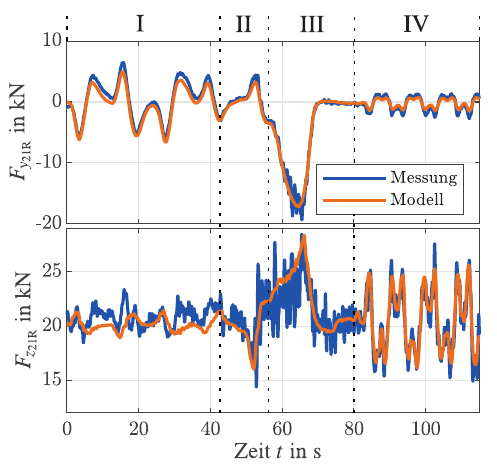}
	\caption{Laterale und vertikale Reifenkraft des rechten Reifens an der ersten Achse des Sattelaufliegers.}
	\label{fig:kräfte_A1R}
\end{figure}

Die simulierten lateralen und vertikalen Reifenkräfte der ersten Achse des Sattelaufliegers werden mit den Messungen der DMS in \textbf{Abbildung \ref{fig:kräfte_A1R}} verglichen. Es ist zu erkennen, dass auch hier das Verhalten der lateralen Reifenkraft in allen Abschnitten \RN{1}-\RN{4} gut abgebildet werden kann (\mbox{RMSE} von $0{,}74\,\textrm{kN}$). Das Verhalten der vertikalen Reifenkraft kann ebenfalls abgebildet werden (\mbox{RMSE} von $1{,}34\,\textrm{kN}$), wobei in Abschnitt \RN{3} eine etwas schlechtere Abbildungsgüte vorhanden ist. 
\begin{figure}[htbp]
	\centering
	\includegraphics[width=\linewidth]{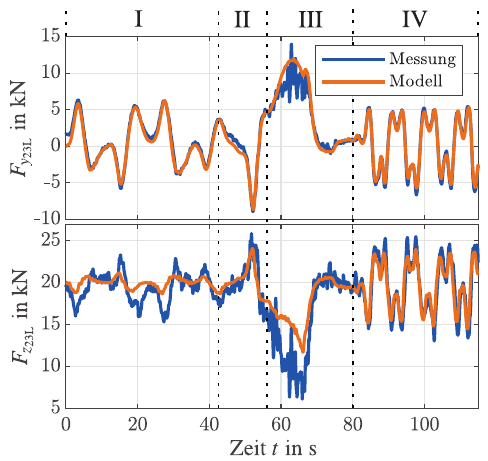}
	\caption{Laterale und vertikale Reifenkraft des linken Reifens an der dritten Achse des Sattelaufliegers.}
	\label{fig:kräfte_A3L}
\end{figure}
Für die Kräfte des linken Reifens der dritten Achse des Sattelaufliegers ist ähnliches zu beobachten (\textbf{Abbildung \ref{fig:kräfte_A3L}}). Das Modell kann die laterale Reifenkraft in Abschnitt \RN{1}, \RN{2} und \RN{4} gut abbilden (RMSE von $0{,}79\,\textrm{kN}$). Lediglich in Abschnitt \RN{3} (enge Kurvenfahrt mit niedriger Geschwindigkeit) kommt es zu einer leicht schlechteren Abbildung des tatsächlichen Verhaltens. Dies ist ebenfalls für die simulierte vertikale Reifenkraft zu beobachten, die in Abschnitt \RN{3} deutlich größer als die gemessene Kraft ist. Der RMSE für die vertikale Reifenkraft beträgt über alle Abschnitte  $1{,}87\,\textrm{kN}$.  Mögliche Erklärungsansätze beziehen sich auf das stark nichtlineare Systemverhalten in diesem Fahrzustand, da für dieses Modell einige Vereinfachungen angenommen wurden, wie lineare Feder- und Dämpfer-Elemente, ein vereinfachtes Reifenmodell (MTFM) und eine Vernachlässigung der Wankabstützung zwischen Zugmaschine und Sattelauflieger an der Sattelplatte. Einen weiteren Einfluss könnte eine leichte Querneigung der Fahrbahn haben. 
Weiterhin tritt bei den Messungen ein erhöhtes Rauschen im Kraftsignal, vor allem in Abschnitt \RN{3}, auf. Dies könnte durch mögliche Stick-Slip-Effekte am Reifen oder durch den zusätzlichen Störeinfluss durch vertikale Straßenunebenheiten hervorgerufen werden.

\section{Zusammenfassung} \label{kap_Zusammenfassung}
In dieser Arbeit wurde ein nichtlineares Zweispurmodell eines Sattelaufliegers vorgestellt, was die Quer- und Wankdynamik abbildet. Zur Modellierung der Reifenquerkräfte wurde eine vereinfachte Variante des Magic Tire Formula Model verwendet. Die unbekannten Modellparameter wurden durch eine Partikel-Schwarm-Optimierung bestimmt. Anhand von der Messung der Gier- und Wankrate sowie den lateralen und vertikalen Reifenkräften an Achse 1 und 3 des Sattelaufliegers wurde die Modellgüte validiert. Trotz einiger vereinfachenden Annahmen, wie lineare Feder-Dämpfer-Elemente oder die Vernachlässigung der Wankabstützung an der Sattelplatte, können die fahrdynamischen Zustände des Sattelaufliegers bei verhältnismäßig geringer Modellkomplexität gut abgebildet werden. Dies legt den Grundstein, um das Modell zur weiteren Entwicklung von Assistenzsystemen, wie der Zustands- und Parameterschätzung mit Beobachtern und Filtern, zu verwenden. Weitere Ansätze zur Steigerung der Modellgüte wären eine detailliertere Modellierung der Feder-Dämpfer-Systeme und der Kopplung zwischen Zugmaschine und Sattelauflieger. \\
Diese Arbeit ist im Rahmen des Projekts \textit{IdenT} (19|19008A, 19|19008B) entstanden. Gefördert vom Bundesministerium für Wirtschaft und Energie aufgrund eines Beschlusses des Deutschen Bundestages.


\clearpage  

\end{document}